\documentclass[12pt]{article}

\usepackage{amssymb}

\begin{document}

\begin{titlepage}

\begin{flushright}
arXiv:0901.1870
\end{flushright}
\vskip 2.5cm

\begin{center}
{\Large \bf Disentangling Forms of Lorentz Violation With\\
Complementary Clock Comparison Experiments}
\end{center}

\vspace{1ex}

\begin{center}
{\large Brett Altschul\footnote{{\tt baltschu@physics.sc.edu}}}

\vspace{5mm}
{\sl Department of Physics and Astronomy} \\
{\sl University of South Carolina} \\
{\sl Columbia, SC 29208} \\

\end{center}

\vspace{2.5ex}

\medskip

\centerline {\bf Abstract}

\bigskip

Atomic clock comparisons provide some of the most precise tests of Lorentz and
CPT symmetries
in the laboratory. With data from multiple such experiments using different nuclei,
it is possible to constrain new regions of the parameter space for Lorentz
violation.
Relativistic effects in the nuclei allow us to disentangle forms of Lorentz violation
which could not be separately measured in purely nonrelativistic experiments.
The disentangled bounds in the neutron sectors are at the $10^{-28}$ GeV level, far
better than could be obtained with any other current technique.

\bigskip

\end{titlepage}

\newpage

There is a great deal of current interest in the possibility that 
Lorentz and CPT invariances may not be exact in nature. Violations of these
symmetries could be tied to quantum gravity, and if any such violation were
observed experimentally, it would be a discovery of profound importance.
Modern tests of Lorentz and CPT symmetry have included studies of matter-antimatter
asymmetries for trapped charged
particles~\cite{ref-bluhm1,ref-gabirelse,ref-dehmelt1} and bound state
systems~\cite{ref-bluhm3,ref-phillips},
determinations of muon properties~\cite{ref-kost8,ref-hughes,ref-bennett1}, analyses
of the behavior of spin-polarized matter~\cite{ref-kost9,ref-heckel2},
Michelson-Morley experiments~\cite{ref-antonini,ref-stanwix,ref-herrmann},
measurements of neutral meson oscillations~\cite{ref-kost10,ref-kost7,ref-hsiung,
ref-abe,ref-link,ref-aubert}, polarization measurements on the light from distant
galaxies~\cite{ref-carroll2,ref-kost11,ref-kost21,ref-kost22},
high-energy astrophysical
tests~\cite{ref-stecker,ref-jacobson1,ref-altschul6,ref-altschul15}, and others.
So far, no significant evidence of Lorentz violation has been found.

There is a parameterization of Lorentz and CPT violations in low-energy effective
field theory, known as the standard model
extension (SME). The SME contains possible Lorentz- and CPT-violating corrections to
the standard model~\cite{ref-kost1,ref-kost2} and general
relativity~\cite{ref-kost12}.
The SME provides a useful framework for interpreting experimental tests of
these symmetries. Many of the coefficients that characterized the SME have been
constrained very tightly. However, many others have not.
(Up-to-date information about bounds on SME coefficients may be found
in~\cite{ref-tables}.)
Moreover, in many cases,
only specific combinations of coefficients can be bounded, rather than the
individual coefficients themselves.

Some of the most precise laboratory tests of Lorentz symmetry are clock comparison
experiments~\cite{ref-berglund,ref-kost6,ref-humphrey,ref-cane,ref-wolf,ref-kornack}.
The results
of these experiments are usually interpreted in the context of the Schmidt
model~\cite{ref-schmidt},
which assigns all of a nucleus's angular momentum to a single unpaired nucleon.
Since these atomic clock experiments are so powerful, there has been a
great deal of interest in devising methods to expand the scope of the bounds they
provide. For example, the suggestion that atomic clock experiments on orbiting
satellites could add sensitivity to previously unconstrained SME
parameters~\cite{ref-bluhm5} attracted a great deal of interest.
In this paper, we shall show that with multiple complementary clock comparison
experiments using different isotopes, there is unexpected sensitivity to new
areas of the SME parameter space.
This fact provides a new motivation for experimenters to improve the atomic clock
constraints on many SME coefficients, even those for which there are already strong
bounds.
(A somewhat similar technique, using complementary experiments with different
materials, was used in~\cite{ref-muller3} to place improved bounds on an
entirely different collection of SME coefficients.)

The minimal SME Lagrange density for a single species of fermion is
\begin{equation}
\label{eq-Lf}
{\cal L}_{f}=\bar{\psi}(i\Gamma^{\mu}\partial_{\mu}-M)\psi,
\end{equation}
where
\begin{eqnarray}
M & = & m+\!\not\!a-\!\not\!b\gamma_{5}+\frac{1}{2}H^{\mu\nu}\sigma_{\mu\nu} \\
\Gamma^{\mu} & = & \gamma^{\mu}+c^{\nu\mu}\gamma_{\nu}-d^{\nu\mu}\gamma_{\nu}
\gamma_{5}+e^{\mu}+if^{\mu}\gamma_{5}+\frac{1}{2}g^{\lambda\nu\mu}
\sigma_{\lambda\nu}.
\end{eqnarray}
Bounds based on atomic clock and other nonrelativistic experiments are
conventionally quoted in terms of specific combinations of SME
coefficients, such as $\tilde{b}_{J}=b_{J}-\frac{1}{2}\epsilon_{JKL}H_{KL}-m
(d_{JT}-\frac{1}{2}\epsilon_{JKL}g_{KLT})$.
(The full set of combinations, which are expressed in sun-centered
coordinates, is given in~\cite{ref-bluhm4,ref-tables}. We shall concentrate on
the $\tilde{b}$ combination, because it is the best measured in most sectors.)
The combinations
mix coefficients from $M$ with mass dimension 1 with coefficients from
$\Gamma$, which are dimensionless and so must be multiplied by $m$ to have
the dimensions match.

In strictly nonrelativistic experiments, the separate coefficients cannot be
further disentangled. Despite this, a number of coefficients can in fact
be separated, using only existing data from laboratory experiments. (The newly
bounded coefficients are not the same as the $\tilde{b}^{*}_{J}$
coefficients, which were introduced in~\cite{ref-tables} and are the analogues of
the $\tilde{b}_{J}$ coefficients for experiments performed with anti-fermions.) What
makes the new bounds possible is a small relativistic effect. Because the effect is
weak, there is a loss of precision in the resulting bounds, compared with
the bounds on the conventional combinations. Nevertheless, these bounds
are important, because they represent separate constraints on groups of
coefficients that it was not previously believed possible to separate.

Nonrelativistically, the effects of, say, $b_{j}$ and $d_{j0}$ are
equivalent,
because $d_{j0}$ enters the Lagrange density in the form
$-\bar{\psi}(Ed_{j0})\gamma^{j}\gamma_{5}\psi=
-\bar{\psi}(md_{j0})\gamma^{j}\gamma_{5}\psi$, which has the same structure
as the $b$ term in ${\cal L}$. However, the nonrelativistic approximation
$E=m$ never holds exactly. In particular, a proton or neutron inside a
nucleus has an energy smaller than its mass, because of nuclear
binding. If the dominant contribution to the measured Lorentz-violating
effects in a atomic species comes from a single unpaired nucleon moving in the
scalar potential of the nucleus,
and the binding energy associated with this nucleon is $e$, what
will be bounded is not $\tilde{b}_{j}$, but rather
\begin{equation}
\label{eq-bdiam}
\stackrel{{\scriptscriptstyle \!\!\!\diamond}}{b_{J}}(e)=
b_{J}-\frac{1}{2}(1+e_{K}/3m)\epsilon_{JKL}H_{KL}-(m-e-2e_{K}/3)d_{JT}+
\frac{1}{2}(m-e-e_{K})\epsilon_{JKL}g_{KLT}.
\end{equation}
$e_{K}$ is the nucleon kinetic energy, and both $e$ and $e_{K}$ are assumed to be
small compared with $m$. We shall neglect $e_{K}$,
although its effects are not that strongly suppressed compared with those of
$e$. With more data, $e_{K}$ could be used to further disentangle $H$ and $g$.

The $\tilde{b}_{J}$ coefficients are the combinations of SME parameters that
appear in the nonrelativistic Hamiltonian for the fermions~\cite{ref-kost26}, which
can be derived from the relativistic theory by a Foldy-Wouthuysen
transformation~\cite{ref-foldy}. However, this transformation process changes in the
presence of interactions. If one instead considers fermions moving in a potential
well, different effective coefficients such as
$\stackrel{{\scriptscriptstyle \!\!\!\diamond}}{b_{J}}$ will appear.

This makes it possible to distinguish contributions from $M$ terms from
those of $\Gamma$ terms. If two experiments are done, using nucleons with
binding energies $e_{1}$ and $e_{2}$, and they produce identical bounds,
conventionally expressed as $|\tilde{b}_{J}|<A$, this actually implies separate
bounds
$|b_{J}-\frac{1}{2}\epsilon_{JKL}H_{KL}|<\frac{2Am}{|e_{1}-e_{2}|}$ 
and
$|d_{JT}-\frac{1}{2}\epsilon_{JKL}g_{KLT}|<\frac{2A}{|e_{1}-e_{2}|}$.
Because the nuclear binding makes a small contribution to the overall
energies of the nucleons involved, the bounds are worse than the
conventionally quoted bounds on $\tilde{b}_{J}$ by a factor of
${\cal O}\left(m/|e_{1}-e_{2}|\right)$. On the other hand, in
experiments involving relativistic motion, in which energies are large
compared with the particle mass, the effects of $M$ and $\Gamma$ terms can
be disentangled without any loss of precision. For example, the maximum achievable
velocity for a particle (which need not be the speed of light when there is Lorentz
violation) generally depends on the $\Gamma$ terms but
not on $M$. However, truly relativistic experiments
have not been nearly as precise as atomic clocks experiments. The most precise
relativistic Lorentz test, which disentangled several contributions to the muon
$\tilde{b}^{\mu}_{J}$, gave bounds at the $10^{-24}$ GeV level~\cite{ref-bennett1}.

As noted in~\cite{ref-kost6}, many
clock experiments will have sensitivities to additional
SME coefficients, beyond those indicated by the Schmidt model. To understand the
additional sensitivities, an improved understanding of nuclear structures would be
required. However, this is a complementary effect to the one considered here. Any
further sensitivities of the type discussed in~\cite{ref-kost6} would still be
sensitivities to the conventional combinations of coefficients and would not allow
the disentanglement of $M$ and $\Gamma$ terms.

The two best atomic clock experiments which measured neutron
Lorentz violations involved $^3$He co-located with heavier species.
The heavy species were used as co-mag\-ne\-to\-me\-ters,
while the $^3$He Zeeman frequencies were read out.
The experiments were ultimately sensitive to the combinations
$\delta\nu_{{\rm He}}-\frac{\mu_{{\rm He}}}{\mu_{M}}\delta\nu_{M}$, where the
$\delta\nu$ and $\mu$ denote frequency shifts and magnetic moments, and the subscript
$M$ indicates the co-magnetometer species.

One experiment used a $^{129}$Xe/$^{3}$He maser.
Taking into account the distinct
binding energies $e_{{\rm He}}$ and $e_{{\rm Xe}}$ of the neutrons in the isotopes
involved, the measurements reported in~\cite{ref-cane} become
\begin{eqnarray}
\label{eq-bnXmixed}
\left|\frac{\mu_{{\rm He}}}{\mu_{{\rm He}}-\mu_{{\rm Xe}}}
\stackrel{{\scriptscriptstyle \!\!\!\!\!\diamond}}{b^{n}_{X}}
(e_{{\rm Xe}})-\frac{\mu_{{\rm Xe}}}{\mu_{{\rm He}}-\mu_{{\rm Xe}}}
\stackrel{{\scriptscriptstyle \!\!\!\!\!\diamond}}{b^{n}_{X}}(e_{{\rm He}})\right|
& = & (-2.2\pm 7.9)\times 10^{-32}\,{\rm GeV} \\
\label{eq-bnYmixed}
\left|\frac{\mu_{{\rm He}}}{\mu_{{\rm He}}-\mu_{{\rm Xe}}}
\stackrel{{\scriptscriptstyle \!\!\!\!\!\diamond}}{b^{n}_{Y}}
(e_{{\rm Xe}})-\frac{\mu_{{\rm Xe}}}{\mu_{{\rm He}}-\mu_{{\rm Xe}}}
\stackrel{{\scriptscriptstyle \!\!\!\!\!\diamond}}{b^{n}_{Y}}(e_{{\rm He}})\right|
& = & (8.0\pm 9.5)\times 10^{-32}\,{\rm GeV},
\end{eqnarray}
neglecting small contributions from other coefficients.

The other experiment used K for the co-magnetometer. The magnetic moment of
K is $\mu_{{\rm K}}\approx\mu_{B}$, three orders of magnitude
greater than $\mu_{{\rm He}}$, so the sensitivity to Lorentz violation
comes almost entirely from the $^{3}$He, provided there are no contributions from the
K valance electron. But the relevant type of electron Lorentz violation is ruled out
very strongly by torsion pendulum experiments~\cite{ref-heckel2}.
So the results from this experiment were~\cite{ref-kornack}
\begin{eqnarray}
\label{eq-bnX}
\left|\stackrel{{\scriptscriptstyle \!\!\!\!\!\diamond}}{b^{n}_{X}}(e_{{\rm He}})
\right|
& = & (-3.7\pm 8.1)\times 10^{-32}\,{\rm GeV} \\
\label{eq-bnY}
\left|\stackrel{{\scriptscriptstyle \!\!\!\!\!\diamond}}{b^{n}_{Y}}(e_{{\rm He}})
\right|
& = & (-9.0\pm 7.5)\times 10^{-32}\,{\rm GeV}.
\end{eqnarray}
Each of the four measurements (\ref{eq-bnXmixed}--\ref{eq-bnY}) implies
a $1\sigma$ bound at the $2\times10^{-31}$ GeV level or better.

In setting these bounds, numerous approximations have been made, and the bounds cannot
be expected to have better than order of magnitude accuracy. The Schmidt model is
a significant idealization, and essentially all nuclear interaction effects
except the binding energy have been ignored.
In reality, the nuclear angular momentum is divided up among protons and neutrons with
different binding energies. The correct value of $e$ must be a weighted average of
these---and this assumes that it is possible to assign
a binding energy to each individual nucleon, which is not strictly the case. The
neglect of $e_{K}$ is also a significant approximation.

However, assuming there are no special cancellations between
the $\stackrel{{\scriptscriptstyle \!\!\!\diamond}}{b_{J}}$ coefficients for
different species, the $2\times10^{-31}$ GeV bound noted above should be
approximately valid. One might be further tempted to conclude that the individual
$b$, $H$, $d$, and $g$ coefficients that make up $\tilde{b}$ should also be bounded
at roughly this level.
There is a crucial difference, however, between possible cancellations among the
Lorentz violation coefficients for different species and between the various
coefficients that make up combinations such as $\tilde{b}^{n}_{J}$. The relative
sizes of the contributions to an observable frequency shift made by the SME
coefficients for different particles are determined by the complex structures of
atoms and nuclei. Even if the various SME coefficients were all comparable in size,
it would be remarkable if their contributions cancelled out. On the other hand,
the relative sizes of the $M$ and $\Gamma$ coefficients that appear in
$\tilde{b}_{J}$ are set by the unknown underlying physics that controls the Lorentz
violation, and it would not be unreasonable for a cancellation to exist within
$\tilde{b}_{J}$.

For example, consider a model in which a single dimensionless vector field acquires a
Lorentz-violating expectation value $u^{\mu}$. (This is known as a ``bumblebee''
model.) The $b$ and $d$
coefficients for a given species of fermions could then be $b^{\mu}=Gmu^{\mu}$ and
$d^{\nu\mu}=Gu^{\mu}u^{\nu}$, where $G$ is a coupling constant that controls the
strength of the interaction between the fermion field and the background. This
coupling would generally depend on the species. So there would
be a natural relationship between the $b$ vectors for different species---they would
all be parallel---but their magnitudes could differ substantially. However, if
the expectation value $u^{\mu}$ is a null vector and $u^{\mu}=(1,\hat{u})$ in a frame
(such as the rest frame of the cosmic microwave background) that is moving
nonrelativistically with respect to the sun-centered frame, $b_{J}-md_{JT}\approx0$.
This demonstrates that cancellation among the coefficients that make up
$\tilde{b}_{J}$ is an entirely natural possibility.

With this in mind, (\ref{eq-bnXmixed}--\ref{eq-bnY}) imply
\begin{equation}
\left|b^{n}_{J}-\frac{1}{2}\epsilon_{JKL}H^{n}_{KL}\right|,
\left|md^{n}_{JT}-\frac{1}{2}\epsilon_{JKL}mg^{n}_{KLT}\right|
< \left(\frac{2m}{e_{{\rm Xe}}-e_{{\rm He}}}\right)
(2\times 10^{-31}\,{\rm GeV})
\end{equation}
for $J=X,Y$. With more data, and including $e_{K}$, we could expect to bound $b$,
$H$, $d$, and $g$ separately at nearly the same level.
%
It remains to determine the difference in the binding energies of the $^{3}$He and
$^{129}$Xe neutrons. $^{3}$He is a relatively compact nucleus, with all three
nucleons in primarily 1S orbital states. The binding energy of 6.69 MeV is nearly
evenly distributed among the constituents, giving $e_{{\rm He}}=2.23$ MeV.
The case of $^{129}$Xe is slightly trickier. This isotope has an average binding
energy of 8.21 MeV per nucleon. However, the neutron whose spin properties the
experiment can observe is a valance neutron and hence less bound than this average
would suggest. As a conservative order of magnitude estimate, we shall therefore take
$e_{{\rm Xe}}-e_{{\rm He}}\approx 4$ MeV.
This yields the final bounds
\begin{equation}
\label{eq-finalbound}
\left|b^{n}_{J}-\frac{1}{2}\epsilon_{JKL}H^{n}_{KL}\right|,
\left|md^{n}_{JT}-\frac{1}{2}\epsilon_{JKL}mg^{n}_{KLT}\right|
<10^{-28}\,{\rm GeV}.
\end{equation}
The strength of these bounds is principally determined by the tightness of the
experimental bounds on the
$\stackrel{{\scriptscriptstyle \!\!\!\!\!\diamond}}{b^{n}_{J}}$, since the
difference in binding energies is close to optimal.
$^{3}$He is a rather weakly bound nucleus, while $^{129}$Xe lies on the broad plateau
of isotopes whose binding energies are close to the maximal value of $\sim 9$ MeV per
nucleon.

Even if they have only order of magnitude accuracy, these are by far the best bounds
on the neutron $M$ and $\Gamma$
coefficients separately. As already noted, it is possible to measure $d$ separately
from $b$ in experiments with relativistic fermions. Often, this is done by
measuring the relationships between energy, momentum, and velocity for extremely
energetic particles. The scale of the resulting bounds is $\gamma^{-2}$, where
$\gamma$ is the Lorentz factor of the particles involved. The most relativistic
observable particles are cosmic ray protons near the Greisen-Zatsepin-Kuzmin limit,
which have Lorentz factors of $\sim10^{11}$. Consequently, it is impossible with
direct measurements of the proton dispersion relation to place bounds on $d$ with
better than $\sim 10^{-22}$ precision. Neutron measurements are far less precise,
since neutrons, being unstable, are less plentiful at the highest energies. The
present bounds represent an improvement over the cosmic ray limits on $d^{n}$
given in~\cite{ref-altschul19} of fourteen orders of magnitude.

Complementary experimental results may also be used to disentangle coefficients in
the proton sector, but the results are not as good. An analysis of the K/$^{3}$He
experiment that includes the proton polarization in the $^{3}$He nucleus gives
bounds on $\stackrel{{\scriptscriptstyle \!\!\!\!\!\diamond}}{b^{p}_{J}}$
at the $10^{-31}$ GeV level. However, these bounds
are not independent of the neutron bounds from the same experiment. The only other
bounds on this form of proton Lorentz violation are at the $10^{-27}$
GeV level~\cite{ref-kost6,ref-humphrey}.
However, some of these bounds come from a H maser experiment, and data collected with
H is ideal for the kind of comparisons we have considered here, since the H nucleus,
being a free proton, has the least possible binding energy of any isotope.
Experiments with H also tend to be very clean, with only proton and electron
coefficients involved.
Combining the $^{3}$He and H data gives disentangled bounds on the $M$ and $\Gamma$
contributions to $\tilde{b}^{p}_{J}$ at the $10^{-24}$ GeV level, still significantly
better than any astrophysical bound.

Improved H maser experiments, in connection with proton measurements made using
heavier nuclei, could obviously give improved bounds on the proton $M$ and $\Gamma$
coefficients.
For both the proton and neutron, there
are also many combinations of Lorentz violation coefficients beyond $\tilde{b}$
which could be separated into their $M$ and $\Gamma$ parts using complementary
experiments.

In summary, we have derived new bounds on several SME coefficients in the neutron
sector. All the coefficients involved have previously been bounded by atomic clock
experiments, but not individually---only in particular combinations. However, when
multiple measurements are available, differences in the nuclear structure of the
isotopes involved make it possible to disentangle the coefficients, a fact which had
not previously been appreciated. Because a weak relativistic effect in the
nucleus is involved, there is a loss of precision in the disentangled bounds of at
least two orders of magnitude. Nevertheless, the new bounds are at the
$10^{-28}$ GeV level, much better that would be possible with any other present
technique.


\end{document}